\documentclass[12pt]{amsart}

\newcommand{\R}{\mathbb R}

\newcommand{\HH}{\mathcal H}
\newcommand{\eqdef}{\stackrel{\text{def}}{=}}

\begin{document}

\title[Quantum field theory on a manifold]
{Mathematical definition of quantum field theory on a manifold}
\author{A. V. Stoyanovsky}
\email{alexander.stoyanovsky@gmail.com}
\address{Russian State University of Humanities}

\begin{abstract}
We give a mathematical definition of quantum field theory on a manifold, and definition
of quantization of classical field theory given by a variational principle.
\end{abstract}

\maketitle

\centerline{\it To the memory of I. M. Gelfand}

\section{Introduction}

In this note we give a definition of quantum field theory (QFT)
on a space-time being a manifold $M$. Such definition is necessary
for unification of QFT
with general relativity. Our definition is almost directly motivated by the definition
of dynamical evolution on space-like surfaces in QFT
on $M=\R^{3+1}$ given in our previous paper [1]. The only essential difference is that we impose
the additional condition that the Hilbert spaces in question be representations of canonical
commutation relations, if the theory is quantization of a classical field theory.
This condition seems reasonable. Classification of unitary representations of canonical commutation relations
can be found, for example, in the book [2] (in the bosonic case).

\section{Definition of QFT on a manifold}

\subsection{}
Let $M$ be a (pseudo-Riemannian) manifold of dimension $D$, and let $G$ be a Lie group acting on $M$.
By definition, a QFT on $M$ assigns

a) to each (space-like)
closed connected co-oriented hypersurface $C$ in $M$ (of codimension 1, below we call them simply surfaces)
a Hilbert space $\HH_C$, and

b) to each closed co-oriented surface $C$ in $M$ with the connected components $C_1$, $\ldots$, $C_n$ it assigns
the space $\HH_C\eqdef\HH_{C_1}\otimes\ldots\otimes\HH_{C_n}$. Here $\otimes$ means
bounded tensor product of Banach spaces, so that for two Hilbert spaces $\HH_1$ and $\HH_2$,
$\overline\HH_1\otimes\HH_2$ (bar means complex conjugation)
is identified with the space
$Hom(\HH_1,\HH_2)$ of bounded linear operators from $\HH_1$ to $\HH_2$;

c) to each manifold $N$ of the same dimension $D$ with the boundary $\partial N$
and a topological type of smooth mappings $N\to M$
which isomorphically map $\partial N$ to a surface $C$ in a compatible way with co-orientation, it assigns a vector
$\Psi_N\in\HH_C$, so that the following conditions hold.

(i) Change of co-orientation of $C$ corresponds to complex conjugation of $\HH_C$.

(ii) If $N$ is the union of two open submanifolds $N_1$, $N_2$ with the common boundary $C_1$, so that $\partial N_1=
C\sqcup C_1$ and $\partial N_2=C_1\sqcup C'$, then $\Psi_N\in\overline\HH_C\otimes\HH_{C'}$ is obtained from
$$
\Psi_{N_1}\otimes\Psi_{N_2}\in\overline\HH_C\otimes\HH_{C_1}\otimes\overline\HH_{C_1}\otimes\HH_{C'}
$$
by contraction $\overline\HH_C\otimes\HH_{C_1}\otimes\overline\HH_{C_1}\otimes\HH_{C'}\to\overline\HH_{C}\otimes\HH_{C'}$.

{\it Corollary.} If we identify $\overline\HH_C\otimes\HH_{C_1}$ with $Hom(\HH_C,\HH_{C_1})$,
then $\Psi_{N_1}$ is a unitary
operator from $\HH_C$ to $\HH_{C_1}$, and its composition with $\Psi_{N_2}:\HH_{C_1}\to\HH_{C'}$ equals
$\Psi_N:\HH_C\to\HH_{C'}$.

(iii) $\Psi_N$ smoothly depends on $C$; hence the bundle with fiber $\HH_C$ over the infinite dimensional
manifold of surfaces $C$ carries a canonical integrable flat connection $\nabla$.

All these data should be compatible with the action of the group $G$ in the obvious sense.

\subsection{Definition of quantization of a classical field theory}
Consider a $G$-invariant classical field theory on $M$ given by the action functional
\begin{equation}
I=\int L(x,\varphi(x),d\varphi(x))dx,
\end{equation}
where $L$ is the Lagrangian depending on points $x\in M$, fields $\varphi(x)$ (we omit the indices of fields),
and their first derivatives $d\varphi(x)$. Then the Euler--Lagrange equations can be written in the
covariant Hamiltonian form, as it is described, for example, in [3,4]:
\begin{equation}
\frac{\delta\Phi}{\delta x^j(s)}=\{H^j(s),\Phi\},
\end{equation}
where $x(s)=(x^j(s))$ is a parameterization of the surface $C$, $x^j$ are local coordinates on $M$,
$\Phi=\Phi(x^j(\cdot);\varphi(\cdot),\pi(\cdot))$
is a functional of
fields $\varphi(s)$ and canonically conjugate variables $\pi(s)$, which changes together with the surface $x=x(s)$;
$H^j(s)=H^j(x(s),x_{s^k}(s),\varphi(s),\varphi_{s^k}(s),\pi(s))$ are the covariant Hamiltonian densities, and
$\{,\}$ is the standard Poisson bracket. Then a QFT on $M$ depending on a parameter $h\ne 0$
is said to be a {\it quantization} of this
classical field theory if the following additional conditions hold:

 (iv) each space $\HH_C$ corresponding to a connected surface $C$ is an irreducible
 unitary representation (in the sense of [2])
 of the canonical commutation relations between the variables
 $\hat\varphi(s)$, $\hat\pi(s)$:
 \begin{equation}
 [\hat\varphi(s),\hat\varphi(s')]=[\hat\pi(s),\hat\pi(s')]=0,\ \ [\hat\varphi(s),\hat\pi(s')]=ih\delta(s-s'),
 \end{equation}
where $[,]$ is the supercommutator;

(v) Consider the flat integrable connection on the bundle $End(\HH_C)=Hom(\HH_C,\HH_C)$ induced from $\nabla$;
denote it by $\nabla_1$. Then in local coordinates $x^j$ on $M$, and for local parameterizations $x=x(s)$
of surfaces $C$, the connection $\nabla_1$ up to $O(h)$ coincides with the differential operator
\begin{equation}
\begin{aligned}{}
\nabla_{1,\frac{\delta}{\delta x^j(s)}}(A)
&=\frac{\delta}{\delta x^j(s)}A-\frac1{ih}[H^j(\hat\varphi(s),\hat\pi(s)),A] \mod O(h)\\
&\equiv\frac{\delta}{\delta x^j(s)}A-\{H^j(s),A\},
\end{aligned}
\end{equation}
where the operators $\hat\varphi(s),\hat\pi(s)$ are put in the Hamiltonian density in their natural order (note that
the covariant Schrodinger functional differential equation in all standard cases does not contain terms like
$\hat\varphi(s)\hat\pi(s)$ which depend on the order of operators); $A=A(x(\cdot)$; $\hat\varphi(\cdot)$, $\hat\pi(\cdot))$
is a regular expression, i.~e. a polynomial expression of smoothed operators
$\int f(s)\hat\varphi(s)ds$ and $\int g(s)\hat\pi(s)ds$
for some smooth functions $f(s)$, $g(s)$.

(vi) For any smooth density $j(x)$ on $M$ with compact support, called {\it source}, and for each co-orientation
of the surfaces $C$, the connection
\begin{equation}
\nabla_j=\nabla+\frac1{ih}\int_C j(x(s))\hat\varphi(s)
\end{equation}
on the bundle $\HH_C$ is also flat.

The latter condition is necessary for construction of the Green functions $\langle\varphi(x_1)\ldots\varphi(x_n)\rangle$,
as in [1].

\end{document}